\documentclass[10pt,a4paper,twocolumn,english,prl,aps,showpacs,floatfix,groupedaddress,superscriptaddress]{revtex4}
\usepackage{times}
\usepackage[]{fontenc}
\usepackage[latin1]{inputenc}
\usepackage{amsmath}
\usepackage{graphicx}
\usepackage{amssymb}

\makeatletter

\def\1{1\negthickspace{\rm I}}

\usepackage{babel}
\makeatother
\begin{document}

\title{Local Measures of Entanglement and Critical Exponents\\ at Quantum
Phase Transitions}

\author{L. Campos Venuti}

\affiliation{Dipartimento di Fisica, V.le C. Berti-Pichat 6/2, I-40127 Bologna,
Italy}

\affiliation{INFN Sezione di Bologna, V.le C. Berti-Pichat 6/2, I-40127 Bologna,
Italy}

\author{C. Degli Esposti Boschi}

\affiliation{Dipartimento di Fisica, V.le C. Berti-Pichat 6/2, I-40127 Bologna,
Italy}

\affiliation{CNR-INFM, Unità di Bologna, V.le C. Berti-Pichat 6/2, I-40127 Bologna,
Italy}

\author{G. Morandi}

\affiliation{Dipartimento di Fisica, V.le C. Berti-Pichat 6/2, I-40127 Bologna,
Italy}

\affiliation{INFN Sezione di Bologna, V.le C. Berti-Pichat 6/2, I-40127 Bologna,
Italy}

\affiliation{CNR-INFM, Unità di Bologna, V.le C. Berti-Pichat 6/2, I-40127 Bologna,
Italy}

\author{M. Roncaglia}

\affiliation{Dipartimento di Fisica, V.le C. Berti-Pichat 6/2, I-40127 Bologna,
Italy}

\affiliation{INFN Sezione di Bologna, V.le C. Berti-Pichat 6/2, I-40127 Bologna,
Italy}

\affiliation{CNR-INFM, Unità di Bologna, V.le C. Berti-Pichat 6/2, I-40127 Bologna,
Italy}

\author{A. Scaramucci}

\affiliation{Dipartimento di Fisica, V.le C. Berti-Pichat 6/2, I-40127 Bologna,
Italy}

\date{\today}

\begin{abstract}
We discuss on general grounds some local indicators of entanglement,
that have been proposed recently for the study and classification
of quantum phase transitions. In particular, we focus on the capability
of entanglement in detecting quantum critical points and related exponents.
We show that the singularities observed in all local measures of entanglement
are a consequence of the scaling hypothesis. In particular, as every
non-trivial local observable is expected to be singular at criticality,
we single out the most relevant one (in the renormalization group
sense) as the best-suited for finite-size scaling analysis. The proposed
method is checked on a couple of one-dimensional spin systems. The
present analysis shows that the singular behaviour of local measures
of entanglement is fully encompassed in the usual statistical mechanics
framework.
\end{abstract}


\maketitle
The entanglement properties of condensed matter systems have been
recently object of intensive studies \cite{osborne02,vidal03}, especially
close to quantum critical points (QCP) where quantum fluctuations
extend over all length scales. Moreover, the amount of entanglement
in quantum states is a valuable resource that promotes spin systems
as candidates for quantum information devices~\cite{lloyd93,bennett00}. 

Since the seminal studies on the interplay between entanglement and
quantum critical fluctuations in spin $1/2$ models \cite{osborne02,osterloh02},
several works suggested different \emph{local} measures of entanglement
(LME) as new tools to locate QCP's \cite{somma04,vidal04,roscilde04,gu04,zanna02,yang05,anfossi05}.
The term local here is meant for measures which depend on observables
that are local in real space. This has to be contrasted with global
measures of entanglement, e.g. the block entropy \cite{vidal03,Cardy04},
or the so-called localizable entanglement \cite{verstraete04b,campos05},
which are aimed to capture the entanglement involved in many degrees
of freedom.

The picture that has emerged so far seems to be non systematic and
model-dependent. Some of the (local) indicators reach the maximum
value at QCP's \cite{gu04}, while others show a singularity in their
derivatives \cite{anfossi05,osborne02,osterloh02,somma04}. Close
to the transition, the system being more and more correlated, one
expects na\"{\i}vely an increase of entanglement. However it seems
that the maxima observed in single-site entropies have to be ascribed
to a symmetry of the lattice Hamiltonian that does not necessarily
correspond to a QCP. For example, in the 1D Hubbard model the single-site
entropy reaches the maximum possible value at $U=0$ \cite{gu04}.
This is due to the equipartition of the empty, singly-, and doubly-occupied
sites rather than to the Berezinski-Kosterlitz-Thouless (BKT) transition
occurring at that point. In fact, in the anisotropic spin-1 system
discussed below, the equipartition points  do not coincide with the
transition lines which are not marked by any symmetry of the lattice
model \cite{chen03}. 

The onset of non-analyticity in two commonly used entanglement indicators
(concurrence and negativity) was recently proved in \cite{wu04} for
models with two-body interactions. Let us first argue, from a statistical
mechanics point of view, that this result is in fact more general:
as a consequence of the scaling hypothesis \emph{every} local average
displays a singularity at the transition with the exception of accidental
cancellations. In particular as any LME is built upon a given reduced
density matrix, the former will inherit the singularities of the entries
of the latter.

A second order quantum phase transition is characterized by long-ranged
correlation functions and a diverging correlation length $\xi$. Let
the transition be driven by a parameter $g$ such that the Hamiltonian
is

\[
\mathcal{H}(g)=\mathcal{H}_{0}+g\mathcal{V}.\]

At $T=0$ the free energy density reduces to the ground state energy
density which shows a singularity in the second (or higher) derivatives
with respect to $g$: 

\[
\frac{1}{L}\langle\mathcal{H}(g)\rangle=e(g)=e_{\textrm{reg}}(g)+e_{\textrm{sing}}(\xi(g)),\]
 where $\xi\approx|g-g_{c}|^{-\nu}$ is the correlation length, $g_{c}$
is the critical point and $L$ is the number of sites. Note that,
as a consequence of the scaling hypothesis, the singular part of the
energy $e_{{\rm sing}}$ is a universal quantity that depends only
on $\xi$, the relevant length scale close to the critical point.
Hence, $e_{{\rm sing}}$ may be considered quite in general an even
function of $g-g_{c}$ around the critical point. 

Differentiating $e(g)$ with respect to $g$, gives the mean value
$\langle\mathcal{V}\rangle/L$, whose singular part $\mathcal{O}_{g}$
behaves as\begin{equation}
\mathcal{O}_{{\it g}}\approx\textrm{sgn}(g-g_{c})|g-g_{c}|^{\rho}.\label{eq:O_g}\end{equation}
Scaling and dimensional arguments imply that $\rho=\left(d+\zeta\right)\nu-1$
where $d$ is the spatial dimensionality and $\zeta$ is the dynamical
exponent. For the sake of clarity here we set $\zeta=1$ as occurs
in most cases \cite{sachdev99}. For a second order phase transition
$\rho>0$. In particular, if $0<\rho\leq1$ the next derivative will
show a divergence %
\footnote{In general, care must be taken to account for logarithmic singularities
see i.e.~\cite{barber83}%
}\[
\mathcal{C}_{{\it g}}\equiv\frac{\partial^{2}e}{\partial g^{2}}\approx|g-g_{c}|^{\rho-1}.\]
In the case where $g$ is mapped to the temperature $T$ in the related
$\left(d+1\right)$-statistical model, ${\cal \mathcal{C}}_{{\it g}}$
will correspond to the specific heat and $\rho=1-\alpha$ (Josephson's
scaling law). As far as entanglement is concerned, the singular term
$\mathcal{O}_{g}$ appears in every reduced density matrix containing
at least the sites connected by the operator $\mathcal{V}$. Obviously,
modulo accidental cancellations, \emph{any} function (i.e.~entanglement
measures) depending on such density matrix, displays a singularity
with an exponent related to $\rho$. The renormalization group theory
allows us to be even more general: to the extent that a local operator
can be expanded in terms of the scaling operators (permitted by the
symmetries of the Hamiltonian), its average will show a singularity
controlled by the scaling dimension of the most relevant term. 

From an operational point of view, LME's have been employed mainly
to detect the transition point using finite-size data. Following the
previous discussion we can argue that, in a typical situation, the
best suited operator for a finite-size scaling (FSS) analysis is precisely
$\langle\mathcal{V}\rangle$ for the following reasons. First because
it naturally contains the most relevant operator, whose average $\mathcal{O}_{g}$
has the smallest possible critical exponent $\rho$. Second the occurrence
of $\textrm{sgn}(g-g_{c})$ in Eq. (\ref{eq:O_g}) plays an important
role in finding the critical point, in case its location is not known
from analytical arguments. The FSS theory asserts \cite{barber83}
that in a system of length $L$, \begin{equation}
{\cal \mathcal{O}}_{{\it g}}(L)\approx\textrm{sgn}(g-g_{c})L^{-\rho/\nu}\Phi_{\mathcal{O}}(L/\xi),\label{eq:FSS}\end{equation}
where $\Phi_{\mathcal{O}}(z)$ is a universal function which must
behave as $z^{\rho/\nu}$ in order to recover Eq. (\ref{eq:O_g})
in the (off-critical) thermodynamic limit $L\gg\xi$. In the critical
regime $z\to0$, $\Phi_{\mathcal{O}}(z)$ must vanish in order to
avoid jump discontinuities for finite $L$. Notice that the sign of
the microscopic driving parameter $\left(g-g_{c}\right)$ survives
in the FSS for ${\cal \mathcal{O}}_{{\it g}}$. As a consequence,
since ${\cal \mathcal{O}}_{{\it g}}(L)$ is an odd function of $\left(g-g_{c}\right)$,
the curves ${\cal \mathcal{O}}_{{\it g}}(L)$ at two successive values
of $L$ as a function of $g$ cross at a single point $g_{L}^{\ast}$
near $g_{c}$ (see below). In this way, by extrapolating the sequence
$g_{L}^{\ast}$ to $L\rightarrow\infty$ one has a useful method for
detecting numerically the critical point. Surprisingly, to our knowledge
the present method was not considered in the past, in favor of the
so-called phenomenological renormalization group (PRG) method \cite{barber83}.
However the PRG method exploits the scaling of the finite-size gap
which requires the additional calculation of an excited level typically
computed with less accuracy than the ground state. This means that
the computational time is roughly doubled. Another advantage w.r.t.
the PRG method is that we do not have the complication of two crossing
points $g_{L}^{\ast}$. In fact as $\Delta\left(g\right)\sim\xi^{-1}\sim\left|g-g_{c}\right|^{\nu}$
is an even function, the scaled gaps will cross at two values of $g$
for each $L$. 

Once $g_{c}$ is determined, using FSS techniques the critical exponents
$\rho$ and $\nu$ may be extracted simply by estimating $\rho/\nu=d+1-1/\nu$.
In order to find other possible critical exponents, we should perturb
our model with other operators permitted by the symmetries $\mathcal{H}\rightarrow\mathcal{H}+g'\mathcal{V}'$
and repeat the same study near $g'=0$.

In what follows we will illustrate these ideas in two different $d=1$
spin models: \emph{i)} the thoroughly studied, exactly solvable Ising
model in transverse field for which all arguments can be checked analytically
and \emph{ii)} the spin-1 $XXZ$ Heisenberg chain with single-ion
anisotropy for which there are no analytical methods to locate the
different transition lines.

\paragraph{Ising Model in Transverse Field.}

\begin{figure}
\includegraphics[%
  width=7cm,
  height=6.5cm]{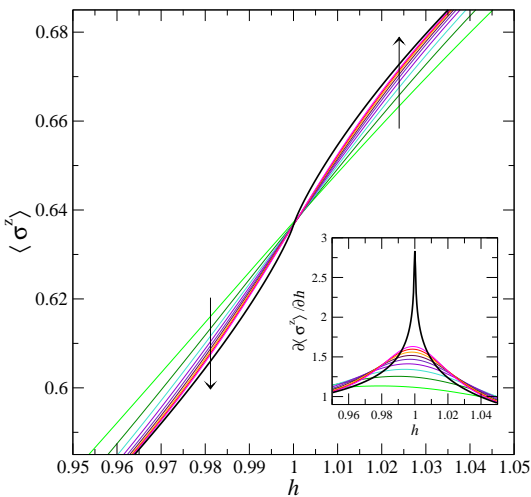}

\caption{The transverse magnetization $\langle\sigma_{i}^{z}\rangle$, is
plotted versus $h$ for various sizes $L$ ranging from 20 to 100
in steps of 10. The black thick line corresponds to the thermodynamic
limit. The arrows indicate the direction of increasing $L$. The inset
shows the derivative w.r.t.~$h$. \label{fig:crossings}}
\end{figure}

We consider the following Hamiltonian with periodic boundary conditions
(PBC)\begin{equation}
H=-\sum_{i=1}^{L}\left[\sigma_{i}^{x}\sigma_{i+1}^{x}+h\sigma_{i}^{z}\right]\,,\label{eq:Ising}\end{equation}
where the $\sigma^{\alpha}$'s are the Pauli matrices. This model
exhibits a QCP at $h=1$, where it belongs to same universality class
as the 2D classical Ising model, with central charge $c=1/2$. From
the exact solution, it is possible to show that the transverse magnetization
$m^{z}=\langle\sigma_{i}^{z}\rangle$, obtained differentiating the
energy w.r.t. the driving parameter $h$, has the following expression
near the transition point $h=1$\begin{eqnarray}
m^{z} & \simeq & \frac{2}{\pi}-\frac{h-1}{\pi}\left(\ln\left|h-1\right|+1-\ln8\right).\label{eq:sz}\end{eqnarray}
As expected, $m^{z}$ is a continuous function at the transition point
$h=1$, showing a singular part which is manifestly odd in $h-1$.
The next $h$-derivative exhibits a logarithmic divergence, as it
is related to the specific heat in the corresponding 2D classical
model. Most important for us is the {}``crossing effect'' near the
critical point of the family of curves $m^{z}(L)$, for different
$L$. In Fig.~\ref{fig:crossings} $m^{z}(L)$ is plotted for several
system sizes. It is evident that, increasing $L$, the crossing points
converge rapidly to $h=1$. 

A quantitative analysis of the crossing effect may be done in the
spirit of FSS, considering separately the critical regime ($L\ll\xi$)
and the off-critical one ($L\gg\xi$) for any finite $L.$ In the
off-critical regime, the finiteness of the correlation length $\xi$,
reflects in the exponential convergence of the energy to the thermodynamic
value \begin{equation}
e_{L}\left(h\right)=e_{\infty}\left(h\right)+\frac{\left|h^{2}-1\right|^{1/2}}{\sqrt{\pi}}\frac{e^{-L/\xi}}{L^{3/2}}\left[1+O\left(L^{-1}\right)\right],\label{eq:en-all}\end{equation}
with $\xi$ given by the formula $\sinh\left(1/2\xi\right)=\left|1-h\right|\left|h\right|^{-1/2}\!\!/2$,
from which we can read the critical exponent $\nu=1$, for $h\to1$
. In the critical regime, the finite-size expression for the transverse
magnetization is\begin{equation}
m_{L}^{z}(h)\simeq\frac{2}{\pi}+\frac{\ln\left(L\right)+\ln\left(8/\pi\right)+\gamma_{C}-1}{\pi}(h-1)+\frac{\pi}{12}\frac{1}{L^{2}}\,,\label{eq:sz_app}\end{equation}
 where $\gamma_{C}=0.5772\dots$ is the Euler-Mascheroni constant.
The finite-size critical field $h_{c,L}$ is obtained via the crossing
points between curves with slightly different lengths, $m_{L}^{z}\left(h_{c,L}\right)=m_{L+2}^{z}\left(h_{c,L}\right)$.
The solution is \[
h_{c,L}=1+\frac{\pi^{2}}{6}\frac{1}{L^{2}}+O\left(L^{-3}\right)\:,\]
showing a convergence towards the critical point as fast as $L^{-2}$. 

As we stressed already, the singularities of local averages reflect
in the behavior of LME's. Among these, the simplest measures the entanglement
between one site and the rest of the system and is given by the von
Neumann entropy $S_{1}=-{\rm Tr}\rho_{1}\ln\rho_{1}$, where $\rho_{1}$
is the reduced single-site density matrix. For spin-1/2 systems $\rho_{1}$
is simply written in terms of Pauli matrices \[
\rho_{1}=\frac{1}{2}\left(\1+m_{x}\sigma^{x}+m_{y}\sigma^{y}+m_{z}\sigma^{z}\right)\,.\]
 For the Ising model (\ref{eq:Ising}) $m_{y}=m_{x}=0$ and the single
site entropy behaves as $S_{1}\sim-0.239\left(h-1\right)\ln\left|h-1\right|$
so that its $h$-derivative diverges logarithmically. A non-zero value
of $m^{x}$ is possible if spontaneous symmetry breaking is taken
into account, by adding a small longitudinal (i.e.~along $x$) field
that tends to zero after the thermodynamic limit is performed. In
this case $\sigma_{i}^{x}$ becomes the most relevant operator and
$m_{x}=\theta\left(1-h\right)\left(1-h^{2}\right)^{1/8}$. Accordingly
the singular part of the entropy is $S_{1}\sim\left(1-h\right)^{1/4}$
for $h<1$. The same singularities are encountered in all the single-site
measures built upon $\left(\rho_{1}\right)^{2}$, e.g.~purity and
linear entropy \cite{somma04}.

On the same line one can consider LME's based on the two-site density
matrix $\rho_{ij}$, obtained taking the partial trace over all sites
except $i$ and $j$. The entries of $\rho_{ij}$ now depend also
on the two-point correlation functions $\langle\sigma_{i}^{\alpha}\sigma_{j}^{\beta}\rangle$.
In accordance with the general theory, all such averages behave as
$\left(h-1\right)\ln\left|h-1\right|$ close to the critical point.
In the case of nearest-neighbour sites this explains the logarithmic
divergence in the first derivative of the concurrence $C(1)$, as
found in \cite{osterloh02}. Instead the leading singularity in the
next-nearest neighbour concurrence $C\left(2\right)=\left[\langle\sigma_{i}^{x}\sigma_{i+2}^{x}\rangle-\langle\sigma_{i}^{y}\sigma_{i+2}^{y}\rangle+\langle\sigma_{i}^{z}\sigma_{i+2}^{z}\rangle-1\right]/2$
turns out to be of the form $\left(h-1\right)^{2}\ln\left|h-1\right|$
\cite{osterloh02}. This is due to the accidental cancellation of
the $\left(h-1\right)\ln\left|h-1\right|$ terms contained in the
correlators.

\paragraph{Spin-1 Heisenberg Chain with Anisotropies. }

\begin{figure}
\includegraphics[%
  width=7cm,
  height=6.5cm]{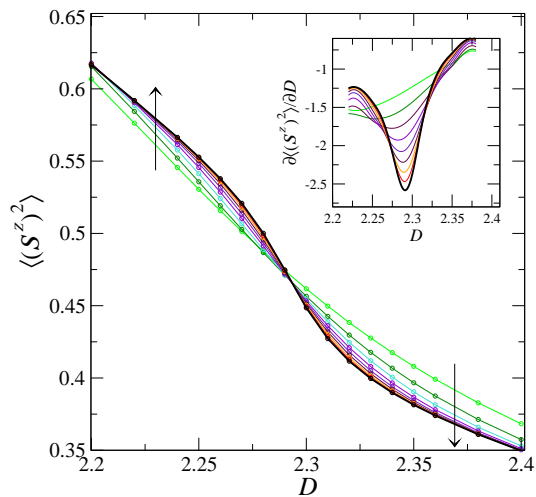}

\caption{The single-site average ${\cal \mathcal{O}}_{D}=\langle\left(S_{i}^{z}\right)^{2}\rangle$,
is plotted versus $D$ for various sizes $L$ ranging from 20 to 100
(thick line) in steps of 10. The arrows indicate the direction of
increasing $L$ and the inset shows numerical derivative w.r.t. $D$,
interpolated with splines. The data have been obtained via a DMRG
program \cite{white93} using $400$ optimized states and 3 finite
system iterations with PBC\label{fig:spin-1}}
\end{figure}

Let us now consider the non-integrable spin-1 model 

\begin{equation}
H=\sum_{i}\left[S_{i}^{x}S_{i+1}^{x}+S_{i}^{y}S_{i+1}^{y}+\lambda S_{i}^{z}S_{i+1}^{z}+D\left(S_{i}^{z}\right)^{2}\right]\label{eq:XXZ1}\end{equation}
which shows a rich phase diagram \cite{chen03}. It is known that
the transition line between the large-$D$ phase (where the spins
tend to lie in the $xy$-plane) and the Haldane phase (characterized
by non zero string order parameters) is described by a conformal field
theory with central charge $c=1$ \cite{cristian03}. This means that
the critical exponents change continuously along the critical line.
For the detection of the $c=1$ critical line, the PRG \cite{glaus84}
or the twisted-boundary method \cite{chen03} have been used in the
literature. We have tested the finite-size crossing method outlined
above, fixing $\lambda=2.59$ for which previous studies ensure $\rho<1$
\cite{cristian03}. The driving parameter being now $D$, the quantity
to consider is $\partial e/\partial D$ which, by translational invariance
reduces to $\langle\left(S_{i}^{z}\right)^{2}\rangle\equiv{\cal \mathcal{O}}_{D}$.
In Fig.~\ref{fig:spin-1} we plot ${\cal \mathcal{O}}_{D}$ versus
$D$ for various sizes $L$. The crossing points of the curves for
subsequent values of $L$, determined by $\mathcal{O}_{D}(D,L)={\cal \mathcal{O}}_{D}(D,L+10)$,
converge rapidly to the critical point $D_{c}=2.294$ consistently
with the phase diagram reported in \cite{chen03}.

The effective theory in the continuum limit of the model (\ref{eq:XXZ1})
around the $c=1$ line reduces to the sine-Gordon Hamiltonian density
\cite{cristian03} 

\begin{equation}
\mathcal{H}_{SG}=\frac{1}{2}\left[\Pi^{2}+\left(\partial_{x}\Phi\right)^{2}\right]-\frac{\mu}{a^{2}}\cos\left(\sqrt{4\pi K}\Phi\right).\label{eq:SG}\end{equation}
The coefficient $\mu$ is zero along the critical line, $a$ is a
short distance cut-off of the order of the lattice spacing and $K$
is related to the compactification radius, varying continuously between
1/2 and 2 along the critical line. In this framework, crossing the
critical line in the lattice model (\ref{eq:XXZ1}) means going from
negative to positive values of $\mu$ and the corresponding $\mu$-derivative
gives ${\cal \mathcal{O}}_{\mu}=\left\langle \cos\left(\sqrt{4\pi K}\Phi\right)\right\rangle $.
From the sine-Gordon theory \cite{lukyanov97} it is known that ${\cal \mathcal{O}}_{\mu}\approx\textrm{sgn}\left(\mu\right)\left|\mu\right|^{K/(2-K)}$
and $\xi\approx\left|\mu\right|^{1/(K-2)}$. In our case $\mu\approx(D-D_{c})$
at fixed $\lambda$, so that $\rho=K/\left(2-K\right),\,\nu=1/\left(2-K\right)$.
On the one hand, the critical exponent $\rho/\nu=K$, can be independently
calculated from the conformal spectrum obtained numerically, as explained
in Ref.~\cite{cristian03} giving $K=0.76$. On the other hand, from
the FSS of the derivatives of ${\cal \mathcal{O}}_{D}$ at $D=D_{c}$
(shown in the inset of Fig.~\ref{fig:spin-1}) we find $K=0.78$
showing that the method presented here is effective for the calculation
of the critical exponent as well. Since the same transition can be
driven by $\lambda$ at fixed $D$, we checked that sitting at $D=2.294$
we obtained $\lambda_{c}=2.591$ by looking at $\mathcal{O}_{\lambda}=\langle S_{i}^{z}S_{i+1}^{z}\rangle$.
According to our general discussion, a similar behavior is seen also
in ${\cal \mathcal{O}}_{D}$ even if it is a single-site indicator.

The scaling exponent $\rho/\nu$ in Eq.~(\ref{eq:FSS}) is best obtained
from the analysis of the first derivative when $\rho<1$. As we move
towards the BKT point, $K\to2$, $\nu\to\infty$, so the divergence
should be seeked in derivatives with increasing order. Accordingly
the crossing method (as well as the PRG) becomes less efficient as
we approach the BKT point, for which a finer analysis is needed involving
level spectroscopy \cite{kitazawa96}. 

Again the singularities of local averages enter the LME's. In the
spin-1 case, thanks to the symmetries of the Hamiltonian (\ref{eq:XXZ1})
the the single-site entropy reads\[
S_{1}=-{\cal {\cal \mathcal{O}}}_{D}\log\left(\frac{{\cal \mathcal{O}}_{D}}{2}\right)-(1-{\cal {\cal \mathcal{O}}}_{D})\log\left(1-{\cal {\cal \mathcal{O}}}_{D}\right),\]
 where $0<\mathcal{O}_{D}<1$ in any bounded region of the phase diagram.
Note that the maximum of $S_{1}$ occurs for ${\cal {\cal \mathcal{O}}}_{D}=2/3$,
which is not related to any phase transition, but simply signals the
equipartition between the three states $|+1\rangle$, $|0\rangle$,
$|-1\rangle$. This occurs for example at the isotropic point ($\lambda=1$,
$D=0$) where the system is known to be gapped. Similarly, in the
Ising model $S_{1}$ is maximal at $h=0$, i.e. when $m_{z}=0$, where
no transition occurs. Therefore the intuitive idea of the local entropy
$S_{1}$ being maximal as a criterion to find quantum phase transitions
\cite{gu04}, seems to be more related to symmetry arguments rather
than to criticality.

In this Letter we have put in evidence the origin of singularities
in LME's which have been recently proposed to detect QCP's. Typically,
apart from accidental cancellations, such singularities can be traced
back to the behavior of the transition-driving term $\mathcal{V}$
and to the corresponding scaling dimension. Moreover the FSS of $\langle\mathcal{V}\rangle$
turns out to be a valuable method to determine the critical point
and the associated exponents. This method has been illustrated for
a couple of spin models displaying qualitatively different QCP's.
More generally, these considerations can be directly transposed to
other many-body problems, like strongly interacting fermionic systems.
Our arguments indicate that the singular behavior of LME's can be
adequately understood in terms of statistical-mechanics concepts.
Physically, the understanding of the intimate relation between genuine
multipartite entanglement and the critical state remains an open challenge.
From this perspective, it may be useful to conceive nonlocal indicators
that could unveil the role of non-classical correlations near criticality.

We are grateful to E. Ercolessi, F. Ortolani and S. Pasini for useful
discussions. This work was supported by the TMR network EUCLID (No.~HPRN-CT-2002-00325),
and the COFIN projects 2002024522\_001 and 2003029498\_013.  

\bibliographystyle{apsrev}
\bibliography{/home/campos/myart/local_entang/entang}

\end{document}